# Creep-fatigue interactions in a polycrystalline structural material under typical high-temperature power plant operating conditions

Markian P. Petkov[a†], Marc Chevalier[b], David Dean[b], Alan C.F. Cocks[a]

[a] Department of Engineering Science, University of Oxford, OX1 3PJ, UK
[b] Assessment Technology Group, EDF Energy, Barnwood, GL4 3RS, UK

Keywords: *crystal plasticity, cyclic hardening, creep-fatigue, micromechanical, structural assessment*

## ABSTRACT

A micromechanical model at the microscale within a crystal plasticity self-consistent model (SCM), is used to analyse loading histories in Type 316H stainless steel, common to structural components in high-temperature power plants. The study compares the SCM predictions on changes in mechanical behaviour and creep properties to analyses via the UK's R5 structural assessment code. Plant-relevant cyclic-creep histories in Type 316H stainless steel at 550°C are examined with focus on the estimation of accumulated creep strain. This aims to quantify how better understanding of creep deformation under cyclic loading can inform R5 code life assessment methods, which are strain based. The effect of cyclic plasticity on primary and secondary creep is quantified. The levels of creep strain accumulated during different dwells, following loading from different macroscopic stress states, is also evaluated and link is made to experimental observations in explaining the response. The impact of displacement- and load-controlled dwells on the evolution of the cyclic hysteresis loop is examined which revealed how different assumptions for creep strain accumulation could employ information from SCM results to reduce uncertainty. It also provides guidance in the development of interpolation frameworks for structural parameters, based on the SCM results which could be used directly in structural assessment and design approaches to estimate the accumulation of inelastic strain with a lower degree of conservatism.

## NOMENCLATURE

| | |
|---|---|
| $C_{ij}$, $E$ | Elastic constants and Young's modulus |
| $E_{ij}$, $E^c$, $\Delta E$ | Macroscopic strain tensor, macroscopic creep strain and macroscopic strain range |
| $j_s$, $j$ | Fitting parameters - hardening |
| $L_{di}$, $L_p$, $L_s$ | Spacing of dislocation junctions, precipitates, solutes |

[†] Corresponding author. Tel.: +44 (0) 1865 283245
Email address M. Petkov: markian.petkov@eng.ox.ac.uk
A.C.F. Cocks: alan.cocks@eng.ox.ac.uk
M. Chevalier: marc.chevalier@edf-energy.com
D. Dean: david.dean@edf-energy.com

| | |
|---|---|
| $\Delta L_r$ | Annihilated dislocation segment length |
| $N_{di}, N_0$ | Current and initial number density of dislocation junctions |
| $p, q$ | Flow rule exponents |
| $T, T_m$ | Temperature, melting temperature |
| $v$ | Poisson's ratio |
| $W_c$ | Fitting parameter – static recovery |
| $\alpha_0$ | Obstacle bypassing strength |
| $\dot{\gamma}^{(\alpha)}, \dot{\gamma}_0$ | Resolved strain rate and reference shear strain rate |
| $\Delta\varepsilon$ | Strain increment |
| $\Sigma_{ij}, \Sigma_{sat}, \Sigma_{dwell}$ | Macroscopic stress tensor, saturated stress, dwell stress or initial stress |
| $\tau, \tau_{cr}, \bar{\tau}_m^r$ | Resolved and critical resolved shear stress; Type III residual stress |
| $\tau_d, \tau_p, \tau_s$ | Internal resistance - dislocation, precipitates, solutes |

## 1. INTRODUCTION

Critical components in high-temperature power generating reactors, e.g. the UK-operated Advanced Gas Cooled Nuclear Reactors (AGRs) and new Gen IV reactor designs, operate under temperatures and loading histories where creep-fatigue deformation of the material could become life-limiting. One of the requirements is a deeper understanding of the interactions between cyclic and creep deformation in austenitic Type 316H stainless steel, typically found in AGR boiler section, under complex thermo-mechanical loading histories that give rise to both primary and secondary loads. Note that under AGR operating conditions (500-600°C), structural failure under strain-controlled cyclic loading histories with displacement- or load-controlled creep dwells is dominated by creep degradation processes. The R5 [1] structural integrity assessment procedure, employed by EDF Energy to assess creep dominated failures, is based on a ductility exhaustion approach [2]. The application of this structural assessment code requires the accurate evaluation of the accumulated inelastic strains during the class of cyclic deformation histories described above. Often, simple material models are employed and assumptions are made about the extrapolation of creep data from laboratory to long-term reactor operating conditions. This can lead to excessive conservatism, since important features concerning the interaction between short-term plasticity and long-term creep behaviour are not modelled explicitly.

There are a number of key experimental observations on austenitic stainless steels that any model needs to replicate if it is to be employed to model creep and relaxation under typical plant loading conditions. Investigations of creep-fatigue interactions in austenitic stainless steel demonstrate a significant decrease in the creep strain accumulation during dwells after plastic pre-straining, which is found to be as a result of cyclic hardening of the microstructure [3,4]. Kikuchi and Ilschner [3] observe reductions in both primary and secondary creep strain rates. On the other hand, Ajaja and Ardell [5] and Fookes et al. [6] demonstrate that prior plasticity can have subtle or no influence on the minimum creep rate in austenitic stainless steels, which is in contradiction to the findings in [3]. The presence of stress relaxation dwells during cyclic deformation gives rise to greater inelastic strain [7]. In addition, due to the induced softening during the dwell, the saturated stress amplitude of the hysteresis loop decreases, compared to that for rapid cycling [7]. The current structural assessment codes recognize the modification of the hysteresis loop when dwells are present. However, the R5 procedure is overly-conservative across the temperature range 550-600°C, and non-conservative in the prediction of inelastic strain enhancement [7]. The positioning of the creep dwell in the cycle with respect to strain is of interest to plant operating conditions. In laboratory experiments, dwells at maximum load are typically employed [7]. It has been suggested that a decrease of the saturated stress amplitude is expected when dwells are not positioned at the maximum tensile strain position (e.g. a relaxation dwell at $\Delta E = 0.0\%$ in a strain controlled cycle). Saturation of the hysteresis loop after a smaller number of cycles is also observed experimentally when the cyclic loading is interrupted by creep dwells. Another possible source of uncertainty in the assessment procedure may arise from the fact that constitutive models, employed by industry, are informed by short-term experimental creep-dwell data ($t_{dwell}$<168 hours). This could have implications when employing procedures based on these models to predict damage development in plant conditions, where dwells can be of the order of 1000 hours.

The simplified constitutive models employed in assessment procedures and design codes are unable to fully capture the trends in behaviour described above. Such creep-fatigue interactions in reactor steels have been studied through finite element modelling techniques [8,9]. However, continuum plasticity models are typically employed in these approaches which do not capture fully the deformation response beyond the available experimental data range. Recently, crystal plasticity finite element (CPFE) models have been used to predict the creep strain accumulation in austenitic stainless steels under creep-fatigue describing the link between macroscopic response and evolution of internal residual stress state, e.g. [10]. Phenomenological constitutive models are typically employed at the slip system level of such CPFE models which brings inherit uncertainty in calibration and difficulty to capture the response dependence on microstructure evolution. Detailed micromechanical models are needed to provide more detailed insights into the cyclic-creep behaviour and how this depends on the evolution of material state. Physically-based models can also potentially provide a means of extrapolating laboratory-extracted information to practical long-term reactor operation. We are not aware of any unified physically-based micromechanical models being employed previously in the literature to explain the experimentally-observed differences in creep response after prior cyclic hardening described above, nor the extrapolation of this behaviour to longer operating cycles. Hu and co-workers [11–13] have developed micromechanical models of elastic-plastic, cyclic and creep deformation of Type 316H stainless steel, employed within a self-consistent modelling (SCM) framework. The SCM models the evolution of obstacles to dislocation motion (e.g. forest dislocations, precipitates and solutes) and the evolution of stress state in a polycrystal. The SCM predicts lattice strains and residual stresses at the grain scale, as well as macroscopic deformation under short- and long-term plasticity, cyclic deformation and creep of Type 316H samples (both solution-treated and ex-service material) [13,14]. The aim of the present study is to use this model framework to evaluate the effects of creep dwells during cyclic loading, similar to those experienced in typical power plant components. The evaluation provides insights into potential improvement of the R5 assessment methods, and guidelines for future experimental studies to validate the predicted behaviour. In addition, this article demonstrates how the physically-rich SCM can be used in a systematic manner to provide extrapolation frameworks which could be used directly in structural design and assessment procedures to reduce conservatism in reactor component lifetime prediction.

The following section briefly introduces the self-consistent modelling (SCM) framework employed in the present study. The micromechanical constitutive model is described in the Appendix. In Section 3.1, the SCM is used to examine the effects of prior plasticity on subsequent creep deformation. In Section 3.2, the effects of cyclic loading on the subsequent creep response is evaluated, with a comparison made between predictions using the SCM and a typical assessment procedure approach. The effects of load-controlled dwells on the accumulation of inelastic strain during a thermo-mechanical history, similar to that experienced in a plant boiler section component, is also evaluated in Section 3.2. A fuller discussion of these results and concluding comments are provided in the last two sections.

## 2. BACKGROUND OF MULTI-SCALE SELF-CONSISTENT MODEL

The three-dimensional SCM framework, developed by Hu et al. [11,12,14] and extended by Petkov et al. [15], captures the anisotropic elastic-plastic deformation of each crystal in a polycrystalline aggregate. The grains are treated as spherical anisotropic inclusions embedded in an elastic matrix. This does not only allow the macroscopic deformation of the polycrystal to be determined, but also allows both i) the evolution of the microstructure within individual grains, and ii) the evolution of residual intragranular stresses due to incompatible plastic deformation of the neighboring grains, to be captured. The framework is based on three different scales – microscopic, mesoscopic and macroscopic – and models the evolution of the mean spacing between individual types of obstacles to dislocation motion, which are treated as internal state variables. The broad range of deformation mechanisms considered in the micromechanical model is shown in Figure 1. In previous studies [11,12,14] it was demonstrated that the model is able to accurately predict mesoscopic lattice strains and macroscopic deformation under monotonic elastic-plastic deformation and creep of ex-service (EX) and solution treated (ST) Type 316H samples, as well as simulate cyclic deformation and stress relaxation [15]. The evolution of precipitate and discrete solute atoms due to long-term aging is also incorporated into the framework [16]. However, the focus of the present study is on ex-service (EX) material where the evolution of such microconstituents is negligible, i.e. their spacing is assumed constant. In addition, the present study evaluates the deformation response of Type 316H under AGR boiler section operating conditions. As such, irradiation effects are not considered in order to focus on high-temperature thermal creep. A brief introduction to the self-consistent framework and the constitutive models is given in the Appendix and a detailed description can be found in [11,12,17,18].

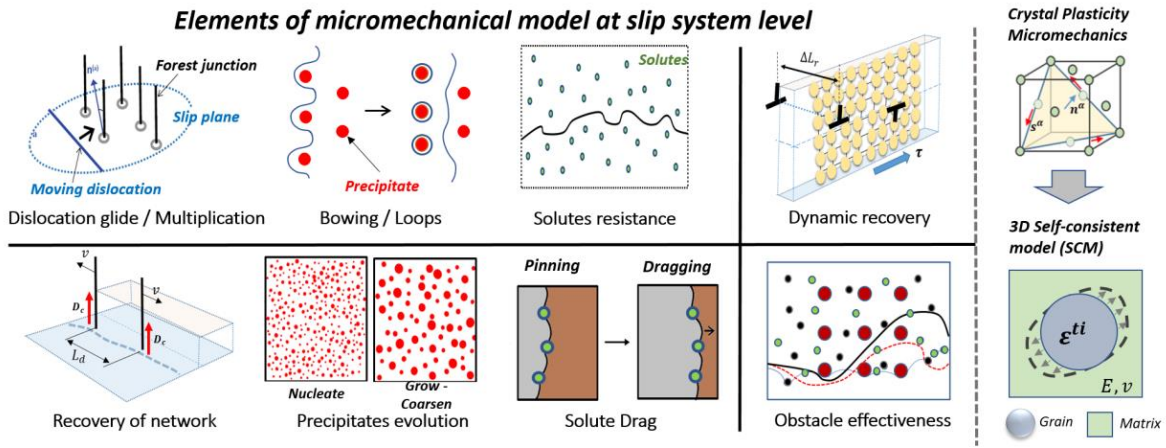

**Figure 1.** Schematic of the SCM modelling framework and the micromechanical deformation mechanisms considered in the constitutive model. The phenomena are related to the evolution of dislocation structures (hardening, static and dynamic recovery), precipitates and solute complexes, as well as the overall effectiveness of obstacles to dislocation motion.

## 3. MODELLING CREEP-CYCLIC DEFORMATION INTERACTIONS

This section employs the multi-scale SCM, introduced in Section 2, to simulate common cyclic-creep deformation histories, relevant to typical high-temperature power plant operating conditions. The model is calibrated to the deformation response of ex-service (EX) austenitic Type 316H stainless steel (Cast 69431) at 550°C, treated as a globally isotropic polycrystalline body (weak texture, see Fig. 2). The chemical composition of the alloy is shown in Table 1. Once calibrated to simple loading histories, as described in the following paragraph, the modelling framework is employed to simulate complex deformation histories, with the resulting predictions being independent of any free parameters. To calibrate the model, the procedure from [15] is adopted using data from four independent experimental tests involving simple uniaxial loading histories: i) elastic-plastic monotonic deformation; ii) displacement-controlled cyclic deformation to saturation of the hysteresis loop; iii) a creep test at constant stress; and iv) a constant-displacement creep (stress relaxation) test following monotonic loading to an initial stress state. Summary of fitting and physical parameters employed in the model here is given in Table 2. The calibrated model predictions of the macroscopic response during creep, relaxation and cyclic deformation are shown in Figure 2. Physical model parameters, such as concentration of solute atoms and spacing of second-phase particles ($M_{23}C_6$ precipitates), were determined from the alloy composition and energy-dispersive X-ray spectroscopy (EDX) studies [13].

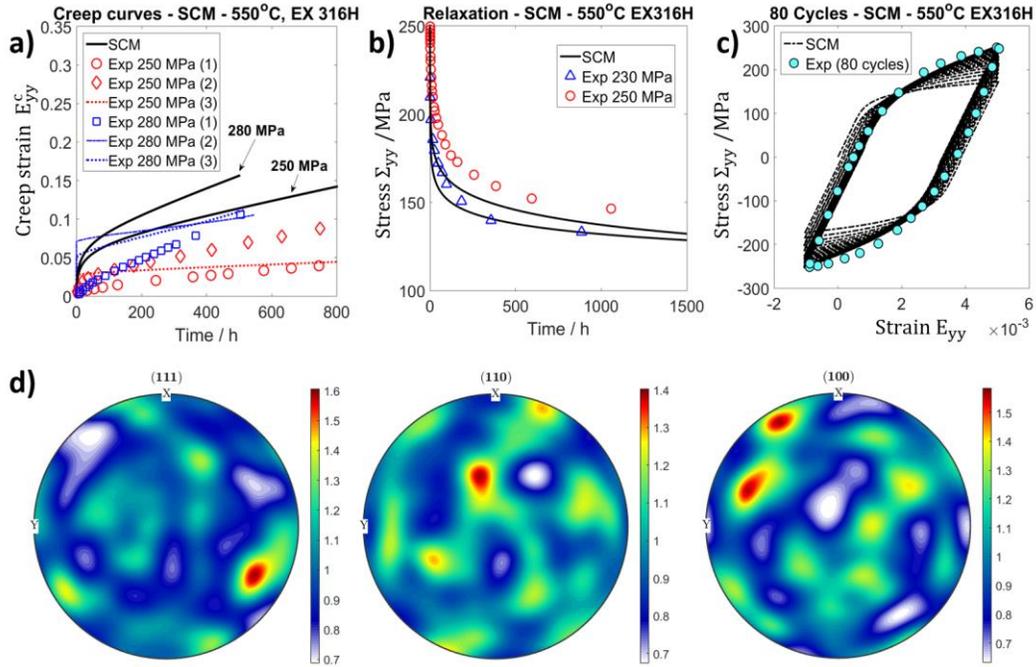

**Figure 2.** SCM predictions and comparison to experimental data on a) creep, b) relaxation and c) cyclic deformation, used to calibrate the modelling framework. Note: $\Sigma$ and $E$ denote macroscopic stress and strain. d) Pole figures of initial texture (100 grains) generated using the MTEX open-source software [19].

**Table 1.** Chemical composition of Type 316H.

| C | Si | Mn | P | S | Cr | Mo | Ni | Co | Fe |
|---|---|---|---|---|---|---|---|---|---|
| 0.07 | 0.41 | 0.98 | 0.024 | 0.017 | 17.1 | 2.3 | 11.6 | 0.07 | Bal. |

**Table 2.** Material parameters for multi-scale model, Type 316H ($T_m$ = 1810 K).

| Parameter | Value | Parameter | Value | Parameter | Value |
|---|---|---|---|---|---|
| $C_{11}$ | 198 GPa* | $N_0$ | $1.5 \times 10^{14}$ | $b$ | $2.5 \times 10^{-10}$ m |
| $C_{12}$ | 125 GPa* | $j_s$ | $2.0 \times 10^{15}$ | $\dot{\gamma}_0$ | 1.0 1/s |
| $C_{44}$ | 122 GPa* | $j$ | $1.0 \times 10^{16}$ | $\alpha_0$ | 0.18 |
| $E$ | 101.3 GPa | $W_C$ | 0.015 | $\tau_p$ | 21 MPa |
| $v$ | 0.39 | $D_C$ | $6.47 \times 10^{-15}$ m$^2$/s ** | $\tau_s$ | 37 MPa |
| $G_0$ | 139 GPa | $\Delta$ | 8 | $\alpha_d$ | 0.35 |

* Data from [17]; ** Parameters from [20].

The extent of latent hardening, captured by the ratio of the hardening parameters introduced in the Appendix as $j_s/j$ was set to 1.4, as discussed in [15]. The hardening parameter $j_s$ and initial value of $N_0$ (i.e. the initial density of dislocation junctions) are computed from the fit to the monotonic elastic-plastic deformation response. The parameter $\Delta L_r \approx \Delta b/\sqrt{3}$, which controls dynamic recovery of the dislocation network (see Appendix), is found from a fit to the cyclic response to saturation for a given number of cycles, using the already determined hardening parameters. The creep and relaxation responses are determined by the parameters $\alpha_0$ for thermally-activated glide and $W_c$ for static recovery. Here, modelling of both creep and relaxation behaviour is considered. A value of 0.18 for $\alpha_0$ provides a good description of the response for a range of materials, as shown in [15]. This value is employed here.

### 3.1 CYCLIC-CREEP DEFORMATION OF TYPE 316H AND THE R5 STRUCTURAL ASSESSMENT PROCEDURE

In this paper, we do not examine various structural assessment procedures or design codes, but rather focus our attention on UK's R5 structural assessment procedure and how the SCM from Section 2 can inform the R5 approach. The current R5 damage assessment procedure [1] uses the stabilized cyclic state stress-range/strain-range response of the material, which is captured by a Ramberg-Osgood model for which the constants are obtained from the AGR Materials Handbook R66 [21]. The R66 data handbook recognizes that softening is induced as a result of the interruption of cyclic histories by creep dwells in each cycle. This type of cycle is representative of deformation histories, experienced during thermo-mechanical loading in the boiler section of high-temperature AGR nuclear power plants in the UK. Introduction of the dwells results in a decrease of the saturated cyclic stress amplitude. Modification of the cyclic curve, following the guidance in [21], is generally conservative in terms of the estimation of start-of-dwell stress. This modification potentially results in an underestimation of the total strain range, and hence, the fatigue damage accumulated. With reference to deformation histories where creep-fatigue failure is dominated by creep, as is the case of AGR plants, the above assumption is conservative. The prediction of a higher start-of-dwell stress overestimates the amount of creep damage, $D_c$, in the dwell via a ductility exhaustion approach ($D_c = \int \dot{\bar{\varepsilon}}_c/\varepsilon_f \, dt$). In addition, the assessment procedures are typically informed by experiments where both the macroscopic strain range (normally $\Delta E > 0.6\%$) and length of the creep dwell ($\sim t_{dwell} = 24$ hours) are different to the actual dwell durations and strain ranges, experienced during plant operation ($t_{dwell} > 1000$ h and $\Delta E < 0.4\%$). Hence, employing the constitutive models from R5/R66 to predict creep strain accumulation in plant with longer dwells introduces further uncertainty in life estimation.

The assessment procedure provides limited advice on the effect of prior cyclic plasticity on subsequent creep response. The conservative approach, which assumes full re-initiation of primary creep ("re-priming") at the start of each dwell, is often adopted. An experimental study by Joseph et al. [4] describes the changes in creep behaviour of Type 316H stainless steel due to prior cyclic plasticity. The authors acknowledge that much of the conducted creep-fatigue tests to examine the effects of prior cyclic deformation on the creep response of the alloy are conducted at large strain ranges (e.g. $\Delta E = 0.8\%$) and short creep dwell times of ~100 hours. The insights from such experimental studies aim to provide guidelines on the modification of material models, used in existing structural assessment procedures and design approaches in industry. However, the experimental configuration of larger strain ranges and short dwell times poses a limitation on the understanding of creep-fatigue interactions in the material under thermo-mechanical histories, typical of plant-operating conditions. In addition, Mamun et al. [22,23] identified a significant difference between the creep strain accumulation during a tensile creep dwell upon re-load from either a compressive or tensile stress state in the cycle. At present, this phenomenon is not accounted for in structural assessment procedures employed in the UK.

The self-consistent model (SCM) from Section 2 could be employed to provide more insights into the accumulation of inelastic strain during complex cyclic loading histories representative of plant operation ($t_{dwell}$ = 1000 hours, $\Delta E < 0.4\%$). The objective is to evaluate the conservatism associated with the modification of the hysteresis loop by reducing the cyclic stress amplitude, proposed in the R66 data handbook. It is also important to understand in greater detail how the creep strain rates during either forward creep (load-controlled) or relaxation (displacement-controlled) dwells evolve with cycles, compared to the assumption that the creep behaviour is not affected by prior cyclic deformation, i.e. full "re-priming" in each cycle. The main aim of the modelling studies on Type 316H in Section 3.1.1 and 3.1.2 is to isolate the effect of prior cyclic hardening on creep deformation, as well as examine the difference in creep strain accumulation during tensile stress dwells following a re-load from either compressive or tensile stress states. Deformation histories in which either load-controlled (forward creep) or displacement-controlled (stress relaxation) dwells are introduced in each cycle are examined in Section 3.2. Dwells in excess of 100 hours are examined. Due to the variety of experimental techniques employed, casts of Type 316H studied and temperature regimes evaluated in the literature [4,24,25], the agreement between model predictions and experimental data is mostly qualitative. This is discussed in Section 4.

**3.2 EFFECTS OF CYCLIC LOADING ON CREEP RESPONSE**

*3.1.1 Evolution of creep strain accumulation with prior cyclic deformation*
The creep-fatigue loading history for ex-service Type 316H at 550°C, simulated in this section, is illustrated in Figure 3 and is described as follows. Initially, a strain range for the cyclic loading test is assumed. In this study, the macroscopic strain ranges $\Delta E$ examined are 0.3%, 0.4% and 0.8%. An example for $\Delta E = 0.8\%$ is shown in Figure 3. The sample is initially loaded in tension to a prescribed strain level, equal to the amplitude of the selected strain range for the cyclic history, shown in Figure 3.a) as position 1. A load-controlled creep dwell is then introduced; the macroscopic start-of-dwell stress ($\Sigma_0$ or $\Sigma_{dwell}$) is recorded. A new sample is then cyclically-deformed for a prescribed number of cycles, e.g. cyclic loading to the third cycle in Figure 3.a), with the cycle ending in a macroscopic compressive state of stress at position 2 in Figure 3.a). The sample is then re-loaded in tension from position 2 to position 3, which is at the same stress as experienced at position 1. A load-controlled creep dwell is then introduced. In Figure 3.b), a similar thermo-mechanical history is illustrated with the cyclic loading ending in a macroscopic tensile stress state at position 4 with the sample then elastically-unloaded to position 5. The sample is then re-loaded in tension to position 6, which is again at the same stress as 1 and 3, where a tensile creep dwell is introduced. The history of Figure 3.b) is evaluated in Section 3.1.2.

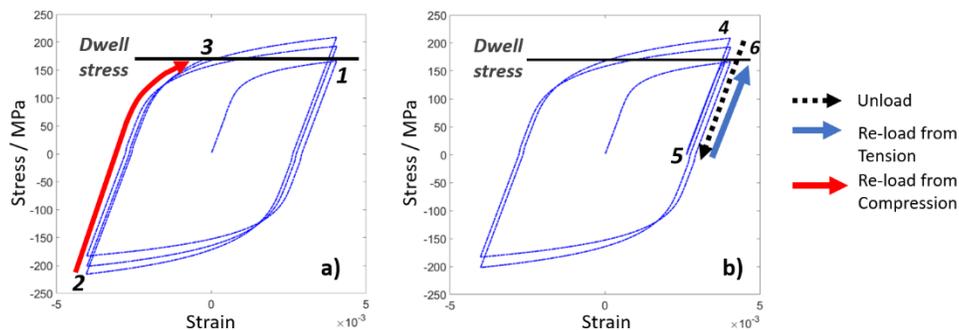

**Figure 3.** Schematic of the deformation history with a tensile stress dwell examined in Section 3.1.1, following a) re-load from compressive and b) re-load from tensile stress state.

An example of the predicted macroscopic creep response following prior cyclic loading after different numbers of cycles for dwells, following reloading from a compressive stress state, as described in Figure 3.a) is shown in Figure 4.a). This figure shows the dwell behavior for $\Delta E = 0.8\%$ and a macroscopic dwell stress of $\Sigma_{dwell} = 167$ MPa. The model predicts a decrease in both the extent of primary creep strain and minimum creep rate with increasing number of cycles to saturation (assumed to be when the change in cyclic stress amplitude between cycles is 1 MPa). Here, the minimum creep rate is arbitrarily selected as the difference in creep strain over the final 20% of creep time. The evolution of macroscopic minimum creep rate $\dot{E}^{ss}$ with increasing $N/N_s$, which is the ratio between cycle number and cycles to saturation of the hysteresis loop, is examined for the three strain ranges in Figure 4.b). We compute and examine the ratio between the minimum creep rate after cyclic loading to a fully-hardened state at the same dwell stress and the minimum creep rate after the initial monotonic load-up - $f_c = \dot{E}^{ss}_{N_s} / \dot{E}^{ss}_{Cycle1}$. The results for $f_c$ are summarized in Table 2. The model predictions for $\Delta E = 0.8\%$ are in agreement with experimental data by Joseph et al. [4] with $f_c \approx 0.4$ for fully-hardened ex-service 316H. At $\Delta E = 0.3\%$, which is more relevant to plant operating conditions, the model predicts $f_c$ of approx. 0.6. This illustrates the diminishing effects of prior cyclic hardening on the subsequent response for shorter strain ranges, which is in qualitative agreement with [4]. The hardening effect is expected to be completely eliminated for strain ranges that fall within the elastic regime.

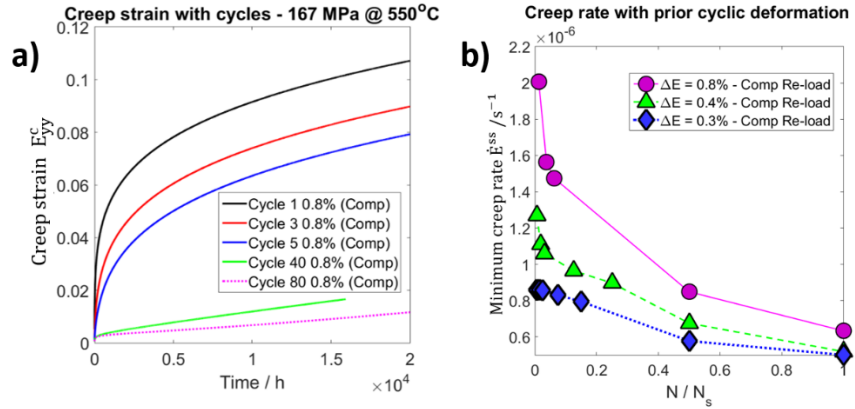

**Figure 4.** a) Model predictions of creep strain accumulation during a dwell following different numbers of prior cycles at $\Delta E = 0.8\%$ with the cycle stopped at point 3 of Figure 3.a). b) Predicted variation of minimum strain rate with $N/N_s$ for the three strain ranges examined.

**Table 3.** Summary of the ratio between the minimum creep rate after cyclic loading to a fully-hardened state at the same dwell stress and the minimum creep rate after initial monotonic load-up (cycle 1) for different $\Delta E$.

| Strain range / Dwell stress | Approx. cycles to Sat - $N_s$ | $f_c$ |
|---|---|---|
| $\Delta E = 0.3\%, \Sigma_{dwell} = 140\ MPa$ | 200 | 0.6 |
| $\Delta E = 0.4\%, \Sigma_{dwell} = 150\ MPa$ | 150 | 0.39 |
| $\Delta E = 0.8\%, \Sigma_{dwell} = 167\ MPa$ | 80 | 0.32 |

*3.1.2 Creep strain accumulation upon re-load from different macroscopic stress states*
The qualitative agreement between the SCM and experimental observations on the effects of prior cyclic hardening on the subsequent creep response, established in Section 3.1.1, provides confidence in the modelling framework in simulating the strain range and creep dwell times of interest. In the present section, the SCM is employed to examine the difference in creep strain accumulation between the cyclic-creep deformation histories, illustrated in Figure 3.a) and Figure 3.b). In Figure 5.a), the accumulation of creep strain during a tensile load-controlled creep dwell after prior cyclic loading to 40 and 80 cycles at $\Delta E = 0.8\%$, following reloading from either a compressive (Figure 3.a) or tensile (Figure 3.b) macroscopic stress state is examined. The results in Figure 5.a) illustrate that during a tensile creep dwell, following reloading from a tensile state of stress, the amount of primary creep strain is far less than the primary creep strain accumulated during the same dwell, following reloading from a compressive state of stress. The effect appears to be more pronounced in fully-hardened samples. The SCM predictions demonstrate a qualitative agreement with observations made by Mamun et al. [26,27] for Type 316H at 650°C. This difference in response is largely related to the different residual stress states generated in the sample during the prior loading history. That generated in the sample after loading from a tensile stress state is closer to that required for steady state creep. Thus, there is less redistribution of stress and less creep strain generated as a result of the stress redistribution process, following reloading from a tensile stress state. This explanation is supported by the observations from [26,27], where this phenomenon is explained by the stress redistribution process between plastically-weaker {220} and plastically-stronger {200},{311} grain families. It also agrees with the observations made by Hu et al. [13] and Hu and Cocks [14] with respect to the effect of stress redistribution on primary creep. It is worth noting that the creep rates for the two different loading types tend to asymptote to the same value during prolonged creep. This is driven by the thermal recovery of the microstructure, represented by the coarsening of the dislocation network, which will tend to lead to a unique material state with time.

Ajaja and Ardell [5] suggest that the effect of prior cyclic hardening on the minimum creep rate during a subsequent dwell is weaker at higher stress. To examine this effect of stress level, another sample is cyclically-loaded at a given $\Delta E$. Again, the cycle ends in either a compressive or tensile macroscopic stress state. At peak hardening, the sample is elastically-unloaded and a tensile creep dwell is carried out at a dwell stress $\Sigma_{dwell} < \Sigma_{sat}$. The ratio between the minimum creep rate for non- and fully-hardened samples from Section 3.1.1, $f_c$, is evaluated in Table 4. The trend in $f_c$ with strain range is found to be similar to that from Table 3 for creep dwells at lower stresses (140-167 MPa). This disagreement between model predictions and findings from [5] suggests the need for further experimental evaluation of the considered histories. Such experiments could either confirm the predicted response or identify missing features in the SCM. Further discussion of this aspect is given in Section 4. Of interest, however, is the deformation behaviour for the plant-relevant strain range $\Delta E = 0.3\%$ and dwell stress $\Sigma_{dwell} = 170$ MPa ($\Sigma_{sat} = 196$ MPa), given in Figure 5.b). The predicted creep response suggests that for shorter strain ranges, relevant to plant operation, the creep response during a tensile dwell, following re-load from a compressive state of stress, resembles closely the creep response during a tensile dwell, following elastic unloading and re-load from a tensile state of stress. Hence, no modifications of the existing models in assessment procedures are required to address the difference in creep strain accumulation during dwells, following reloading from different stress states, if they are to be applied to strain ranges <0.3%.

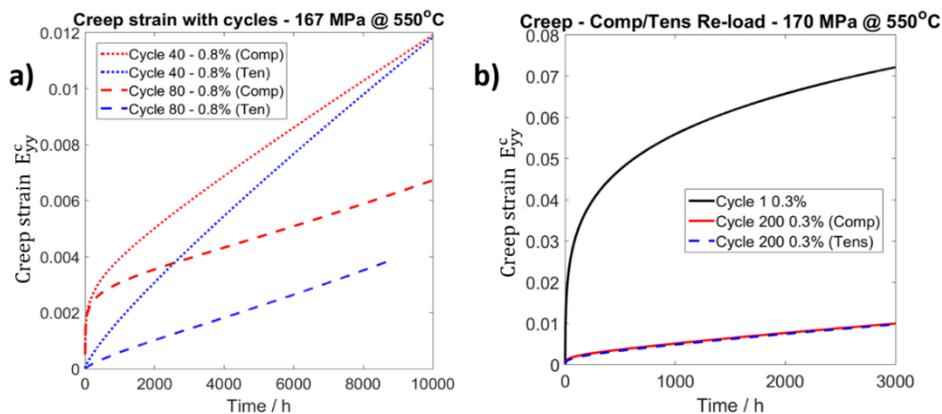

**Figure 5.** a) Differences in creep strain accumulation during a tensile dwell at 167 MPa ($\Delta E = 0.8\%$) after prior cyclic loading to 40 and 80 cycles following reloading from either a tensile or compressive stress state. b) Example of change in creep strain accumulation in the fully-hardened state for $\Delta E = 0.3\%$ and a dwell stress of 170 MPa.

**Table 4.** Summary of the ratio between the minimum creep rate after cyclic loading to a fully-hardened state at the dwell stress ($\Sigma_{dwell} < \Sigma_{sat}$), following reloading from compressive and tensile stress states, and the minimum creep rate after a creep dwell at the same stress for different $\Delta E$. Note: The ratio is equivalent to $f_c$.

| Strain range / Dwell stress | $\Sigma_{sat}$ | $N_s$ | $f_{c,com.re-load}$ | $f_{c,ten.re-load}$ |
|---|---|---|---|---|
| $\Delta E = 0.3\%, \Sigma_{dwell} = 170\ MPa$ | 196 MPa | 200 | 0.75 | 0.75 |
| $\Delta E = 0.4\%, \Sigma_{dwell} = 190\ MPa$ | 214 MPa | 150 | 0.36 | 0.37 |
| $\Delta E = 0.8\%, \Sigma_{dwell} = 240\ MPa$ | 275 MPa | 80 | 0.32 | 0.36 |

### 3.3 EFFECTS OF CREEP DWELLS ON CYCLIC DEFORMATION

#### *3.2.1 Evaluation of cyclic deformation histories with displacement-controlled creep dwells*

In the present section, the SCM is employed to simulate the deformation of ex-service austenitic Type 316H stainless steel, under a cyclic deformation history with displacement-controlled dwells (pure relaxation, no elastic follow-up) introduced at the maximum tensile strain position. The model is calibrated in Section 3.1. A set of experimental data for the deformation of the same cast of material considered above for cyclic loading histories with dwell durations of 1-100 hours at strain ranges $\Delta E$ = 0.6-1.2% is given in [7]. Based on the available experimental data and the plant-relevant strain ranges of the order of $\Delta E < 0.4\%$, the SCM framework is employed here to examine strain ranges which capture both experimental- and operation-relevant strain ranges (i.e. $0.2\% < \Delta E < 0.8\%$) and dwell times varying from 1 to 1000 hours. This systematic study allows direct comparison between the SCM predictions, experimental data and assessment procedure advice via the R66 data handbook on modification of the hysteresis loop. Once agreement is established with the experimental results for dwells of moderate duration (~100 hours), the physically-based SCM could be used to provide deeper understanding during dwells of longer duration.

Examples of the predicted hysteresis loop for $\Delta E$ = 0.3% and 0.8% for relaxation dwells of 24 and 1000 hours are shown in Figure 6 and compared to the pure fatigue hysteresis loops. The modelling results predict the progressive decrease in the saturated cyclic stress amplitude with increasing dwell duration. In addition to the predicted trend in softening of the cyclic response, the modelling results capture the decrease in number of cycles to saturation of the hysteresis loop for longer dwells. This qualitatively agrees with observations by Cronin and Spindler [7] and Sauzay et al. [28]. Examination of the microstructural evolution during the dwell demonstrates an increase in the spacing of dislocation junctions as a result of thermal recovery, in agreement with observations in [15]. This results in fewer obstacles to dislocation motion during the tensile-compressive cycle after the dwell, manifesting itself in the observed softening of the hysteresis loop. In Figure 7.a), predictions of how the saturated stress amplitude varies with inelastic strain (i.e. the sum of creep strain during the dwell and the plastic strain accumulated during the cyclic loading part of the history) are compared with experimental results for $\Delta E$ = 0.6% from [7]. The model correctly captures the dependence between inelastic strain, dwell time and saturated stress amplitude. At $\Delta E$=0.6%, the SCM predicts a value of $\Sigma_{sat}/\max(\Sigma_{sat,no\ dwell}) = 0.8$ for 24-hour dwells in Fig. 8.a), compared to the experimentally-recorded value of ~0.76 for this $t_{dwell}$ and $\Delta E$. More creep strain is accumulated at longer dwells resulting in a larger decrease in the saturated stress amplitude. The saturated stress increases with increasing strain range as a result of the increase in dislocation junctions during the rapid part of the cycle. In addition, the model appears to capture the trend of increasing total inelastic strain with increasing dwell time in the experimental data. These observations provide confidence in the accuracy of this multi-scale physical modelling approach for this type of loading history.

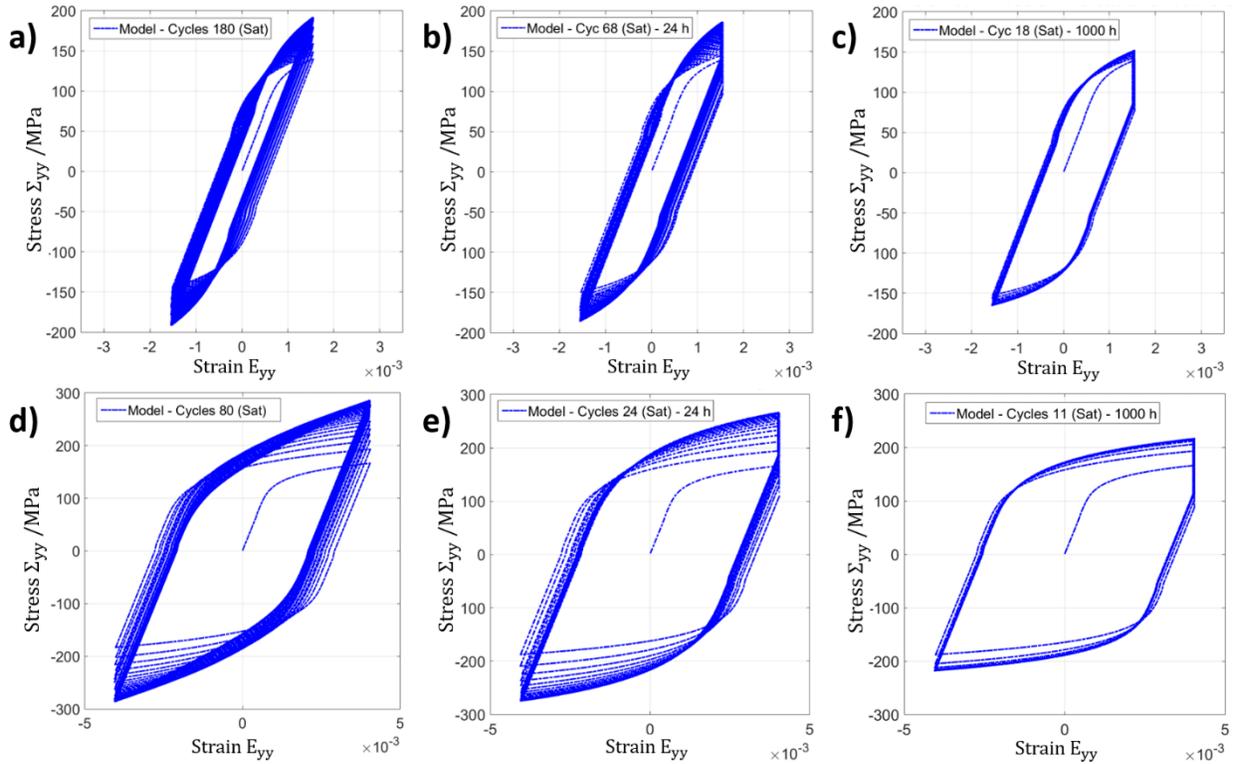

**Figure 6**. Predicted hysteresis loops at $\Delta E = 0.3\%$ for a) pure fatigue and relaxation dwells at maximum tensile strain of b) 24h and c) 1000h. The equivalent predictions for $\Delta E = 0.8\%$ are shown in d)-f).

Comparison between the SCM predictions and R66 handbook advice (which provides a multi-cast experimental data fit) on modification of the saturated stress amplitude with inelastic strain and dwell time is also made in Figure 7.b). Despite the qualitative agreement between the trend of softening of the hysteresis loop predicted by the two approaches, the R66 advice underestimates the softening of saturated stress amplitude significantly. At $\Delta E = 0.3\%$, the predicted decrease in $\Sigma_{sat}$ for a history with 1000-hour dwells from that obtained in a rapid fatigue cycle is ~25 MPa using the R66 data handbook, compared to ~75 MPa by the SCM. An overestimation of the start-of-dwell stress results in an overestimation of the creep strain during the dwell and, therefore, a conservative estimate of the time for failure (or crack initiation) in Type 316H stainless steel. Experimental data on the creep strength of ex-service polycrystalline Type 316H at 550°C, provided by EDF Energy, indicates that failure time decreases by nearly an order of magnitude with a decrease in applied stress of 50 MPa (N.B. Data not shown here). It is worth noting that both experimental results and SCM predictions confirm the lack of conservatism in the R66 advice on the prediction of strain enhancement for cyclic-dwell histories of this type at 550°C.

Comparison between the experimental observations, modelling results and the advice in the R66 data handbook reveals several aspects of existing assessment and design approaches, with emphasis on plant-relevant strain ranges and dwell duration, which can be improved. Based on the consideration that the guidance in R66 is informed by experimental data on short creep dwells, care must be taken to ensure that the practical advice remains conservative when extrapolated to longer dwell times. Based on the predictions of the SCM for long-term dwells (1000 hours) and the general agreement of the SCM with the trend in experimental data for short and moderately long dwells (N.B. no experimental data for 1000-hour dwells is available), the modelling predictions suggest that the R66 prediction will also be conservative at longer dwell times. The results of this section illustrate that the SCM predictions could be used systematically to construct an extrapolation framework for the decrease in saturated stress amplitude with dwell time for different $\Delta E$, using the pure fatigue saturated stress amplitude as a reference point. Informing the assessment advice from this extrapolation framework could reduce the degree of conservatism in estimation of the start-of-dwell stress. Further discussion can be found in Section 4.

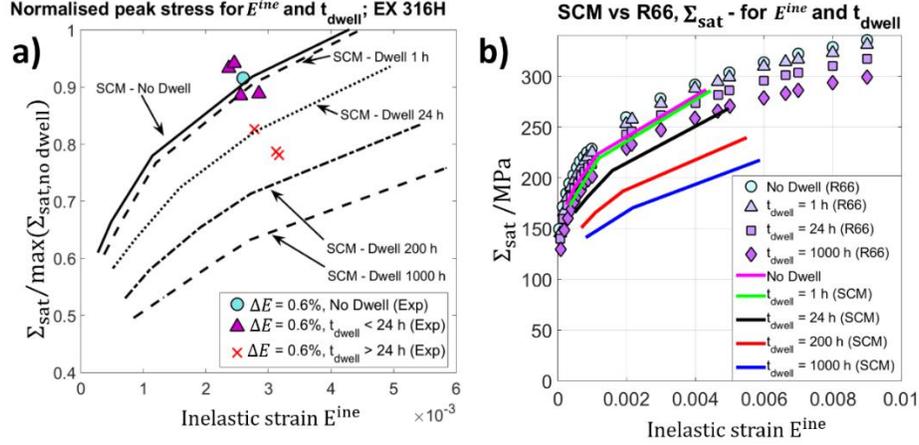

**Figure 7.** a) Comparison between experimental results (Cast 69431) on $\Sigma_{sat}$ with $E^{ine}$ at $\Delta E = 0.6\%$ and SCM predictions. b) Comparison between R66 and SCM predictions on variation of $\Sigma_{sat}$ with $E^{ine}$ and dwell time.

Direct improvements could be made in another aspect of the assessment procedure, related to the evolution of creep strain during dwells in different cycles, through application of the SCM framework. Constitutive models in the assessment procedure do not attempt to predict the evolution of creep dwell strain with cycles. The creep strain during a dwell at a given stress and dwell time is assumed to be the same as that during a dwell at the same stress and time following load-up of a virgin sample to the same start-of-dwell stress. The SCM is employed here to systematically examine the creep strain accumulated during dwells at position $E^{\max,tens}$ (which we denote as $E^{c,dwell(\sigma_0,t)}$) where $\sigma_0$ is the start-of-dwell stress and $t_{dwell}$ is the dwell time. The model is then used to predict the creep strain accumulated during a dwell of the same duration, but following monotonic loading of the virgin material to the same start-of-dwell stress. This is designated here as $E^{c,monoton(\sigma_0,t)}$. We then define a numerical factor $f_d = E^{c,dwell(\sigma_0,t)} / E^{c,monoton(\sigma_0,t)}$, which is the ratio of the creep strains predicted from the two simulations. Predicted values of $f_d$ for different dwell times and strain ranges, as a function of cycle number compared to the number of cycles to saturation, are summarised in Figure 8. The predicted variation of $f_d$ in Figure 8 illustrates that for lower strain ranges and shorter dwell times significant differences are expected between the accumulated creep strain during actual dwells to saturation and the creep strain accumulated during equivalent creep dwells at the same stress without prior cyclic deformation. The value of $f_d$ decreases with increasing number of cycles for a short dwell time of 24 hours at 0.3% and 0.4% strain ranges. Hence, the practical approach of using an equivalent dwell without considering the prior history appears to be conservative for short strain ranges by overestimating the creep strain during the dwell. The difference diminishes for longer dwells and larger strain ranges. The model captures this progressive decrease in the difference between the actual response and the practical approach which is believed to be due to the greater tendency in re-initiation of primary creep at larger strain ranges, suggested in [15]. In addition, the minimum creep rate at larger strain ranges (i.e. higher stresses) has been observed experimentally in [6] to be less influenced by prior plasticity. Figure 8 illustrates that at full saturation for the short strain range of 0.3% and dwell time of 1000 hours, the predicted creep strain within a dwell using the practical R66 approach assuming full "re-priming" could be overestimated by approx. 10%. At shorter dwell times this difference increases, e.g. for a 24-hour dwell for a strain range of 0.3% the creep strain could be overestimated by approx. 35% at $N/N_s = 1.0$. This evaluation of equivalent dwells for a range of strain ranges and dwell times suggests that the ratio $f_d$ could be employed in assessment procedures and design codes as a multiplier on the creep strain predicted for a known start-of-dwell stress. This could be useful for shorter strain ranges, relevant to plant operating conditions, with the aim to reduce the conservatism in dwell creep strain estimation.

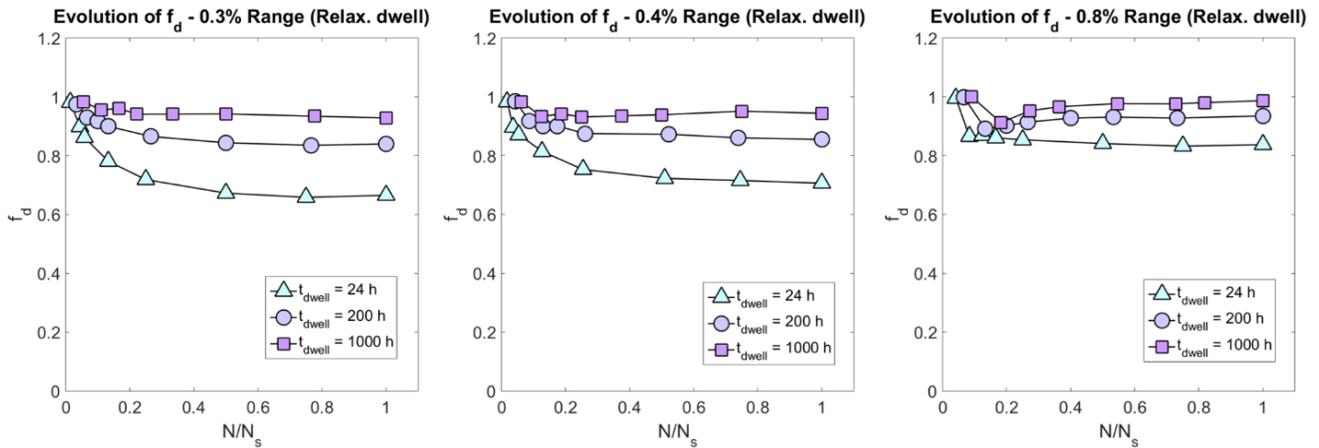

**Figure 8**. Evolution of $f_d$ with total strain range, normalised cycles to saturation and dwell times for relaxation dwells at maximum tensile strain.

The effect of varying the position of the relaxation dwell in the cycle on the saturated stress amplitude of the hysteresis loop is now examined. This cyclic-creep history is highly-relevant to plant operating conditions where dwells may not always be at the maximum tensile strain position. The R5 assessment procedure does not provide any advice on modification of the hysteresis loop with dwell position. The SCM framework is employed here in the same manner, demonstrated in Figure 6. However, the dwell is now positioned at $E = 0.0\%$. Strain ranges between 0.2% and 0.8% and dwell times of 1-1000 hours are examined. The predicted cyclic hysteresis loop of the material at strain ranges of 0.3% and 0.8% for relaxation dwells of 24, 200 and 1000 hours are shown as examples in Figure 9. Faster saturation of the hysteresis loop with longer dwell times is again predicted. However, for dwells at the $E = 0.0\%$ position, the saturation of the cyclic response is reached after a greater number of cycles for a given dwell time as compared to dwells at maximum strain level, as shown in Figure 10.a). Comparison of the predicted variation in saturated stress with inelastic strain with experimental data from [7] on the same ex-service cast of 316H stainless steel for $\Delta E = 0.6\%$ with 24-hour dwells either located at $E = 0.0\%$ or the maximum strain, $E^{max,tens}$ is shown in Figure 10.b). The agreement between SCM predictions and experimental data demonstrates that the softening effect on the saturated stress is less for mid-strain dwells. This is expected due to the lower start-of-dwell stress, confirming the hypothesis from [7]. The softening effect of thermal recovery scales with the spacing of the dislocation junctions (see Eq. A.5), whose value at the start of the dwell is determined by the loading history and the start-of-dwell stress. Although not presented here, it is found that the initial dislocation junction density is higher for dwells at the maximum tensile strain position or peak stress, due to the greater degree of hardening induced, as compared to intermediate dwells. The subsequent rate of thermal recovery during the dwell is therefore lower for intermediate dwells. Hence, the softening effect (or rate of stress relaxation) of intermediate displacement-controlled dwells at lower start-of-dwell stresses on continuing cyclic deformation will be lower as compared to dwells, positioned at peak stress. Similar to observations in Figure 7.a), the model appears to correctly capture the trend of increasing total inelastic strain with increasing dwell time for dwells at both maximum tensile strain and mid-strain positions. This is demonstrated in Figure 10.b). Systematic application of the SCM could quantify the effects of different dwell positions and durations on the decrease of the saturated stress amplitude to construct an extrapolation framework to incorporate the observed trend in assessment methods. This is not further investigated in the present article.

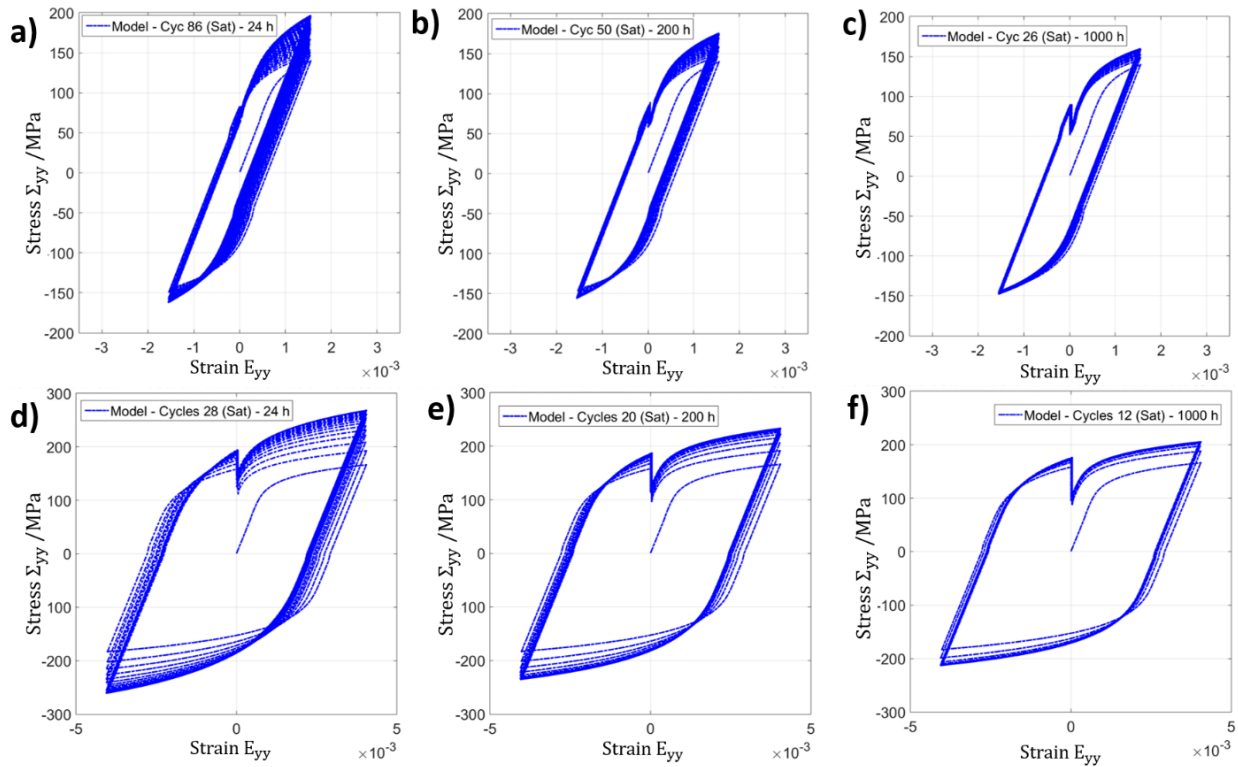

**Figure 9.** Predicted hysteresis loops at $\Delta E$ = 0.3% for a) 24-hour, b) 200-hour and c) 1000-hour relaxation dwells at 0.0% strain position. The equivalent predictions at $\Delta E$ = 0.8% are shown in d)-f).

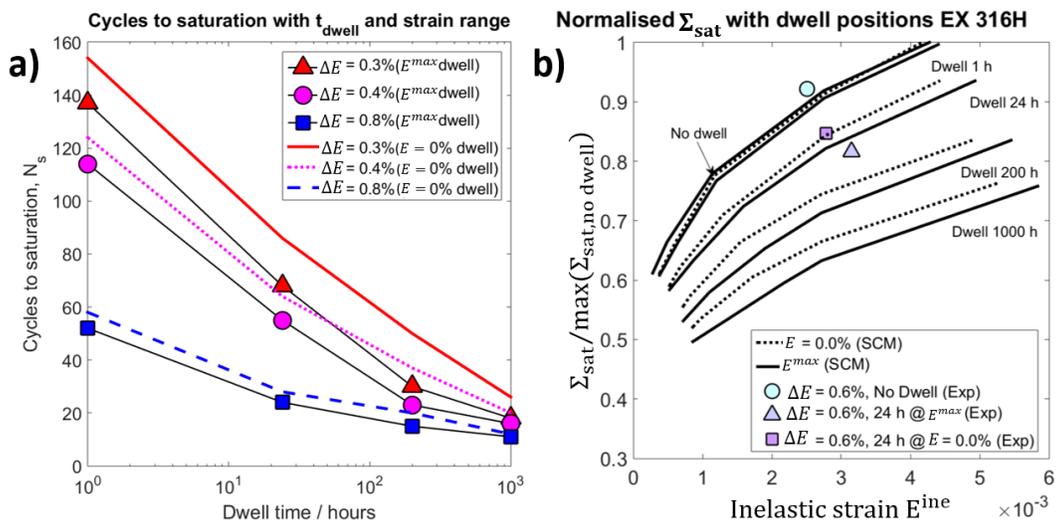

**Figure 10.** a) SCM predictions of cycles to saturation with different cyclic-dwell histories and strain ranges. b) Experimental data on $\Sigma_{sat}$ at $\Delta E$ = 0.6% and SCM predictions for different dwell positions.

*3.2.2 Evaluation of cyclic deformation histories with load-controlled creep dwells*
In this section, the interaction between cyclic and creep deformation is examined during a thermo-mechanical deformation history similar to that experienced by AGR components, such as thin-walled pipes and pressure vessels. In such components, primary loads dominate during the creep dwell. However, plasticity due to cyclic deformation as a result of combined primary and secondary loading can give rise to localized creep-fatigue deformation and subsequent crack initiation [29]. The deformation history examined here comprises of loading to the maximum tensile strain followed by a load-controlled creep dwell, with a constant strain range load reversal at the end of the dwell. The start-of-dwell stress is the same in each cycle. The R5 procedure for damage assessment advises the user to consider two practical approaches for estimation of accumulated creep strain during dwells:

a) Assume that cyclic deformation causes complete removal of primary creep during the dwells (i.e. $E^c$ accumulated during 9 consecutive dwells of 1000 hours is equivalent to $E^c$ accumulated during a single 9000-hour dwell, i.e. no "re-priming");
b) Assume the primary creep stage is fully re-initiated during each dwell (i.e. $E^c$ during nine 1000-hour dwells equal to $9 \times E^c$ during the first 1000-hour dwell).

The SCM is employed in this section to assess which of the two practical assessment approaches provides the most reasonable prediction of the accumulated creep strain during the actual creep-fatigue history as a function of strain range and dwell duration. N.B. Damage initiation is not considered here, as the present study is solely focused on the creep deformation and accumulation of creep strain. The SCM framework is used to simulate the deformation history described at the beginning of this section for strain ranges between 0.3% and 0.8% and load-controlled creep dwells of length 24-1000 hours. Examples for different dwell times at $\Delta E = 0.3\%$ are shown in Figure 11. The creep strain accumulated during the dwells for dwell times of 24 hours, 200 hours and 1000 hours are shown in Figure 12.

Based on the simulated creep response during each dwell, with examples shown in Figure 12, model predictions of cumulative creep strain obtained via the practical approaches a) and b) are compared in Table 5. A cumulative creep dwell duration of 9000 hours is examined. Note that three different dwell durations are considered: namely, 9×1000-hour, 45×200-hour and 375×24-hour dwells. The results in Table 5 illustrate that the difference between the two assessment approaches a) and b), and the simulated history, is strongly dependent on dwell duration and strain range. In all cases examined, approach a) underestimates the accumulated creep strain and remains non-conservative as compared to the simulated dwell history. By contrast, approach b) always remains conservative. It is found that the predicted creep response is much closer to approach a) for the 9×1000h dwell history across all strain ranges with the difference within 10%. For histories with dwells of 24h and 200h, the modelling results demonstrate that approach a) becomes increasingly non-conservative with differences of ~35% for 200-hour dwells at $\Delta E = 0.8\%$, and ~65% and 80% for 24-hour dwells at $\Delta E = 0.3\%$ and 0.8%, respectively. This suggests that approach a) will yield an unreasonable estimation of the accumulated creep strain during dwells shorter than 1000 hours. On the other hand, approach a) estimates the accumulated creep strain for 1000-hour dwells with a reasonable degree of accuracy. This is because the contribution of primary creep diminishes significantly with increasing number of cycles and dwell duration, although it should be noted that a load-reversal will always give rise to a small amount of primary creep [15].

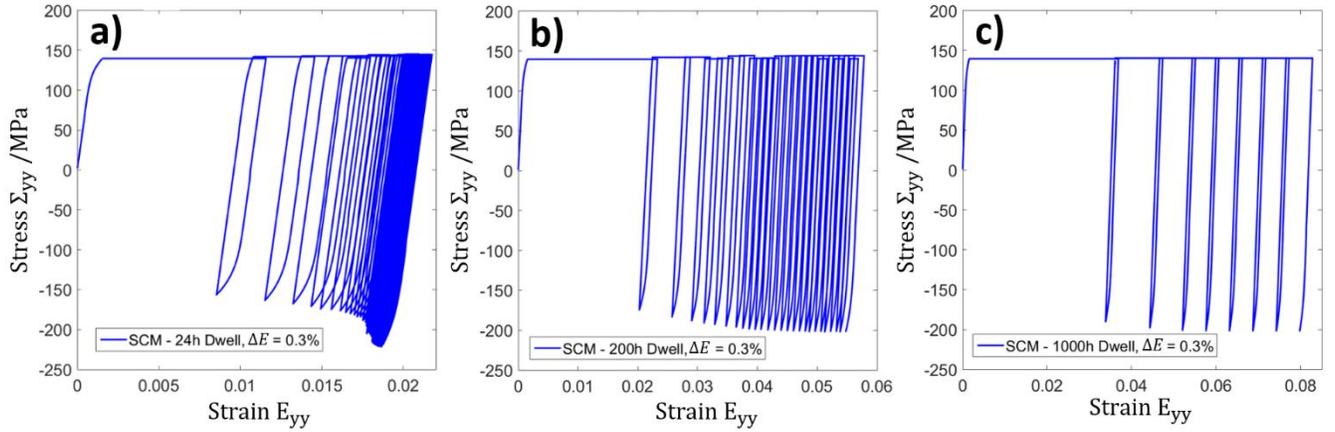

**Figure 11.** Predicted creep-fatigue response at $\Delta E = 0.3\%$ for load-controlled creep dwells of a) 24 h, b) 200 h and c) 1000 h. Note: Macroscopic dwell stress $\Sigma_{dwell} = 140$ MPa.

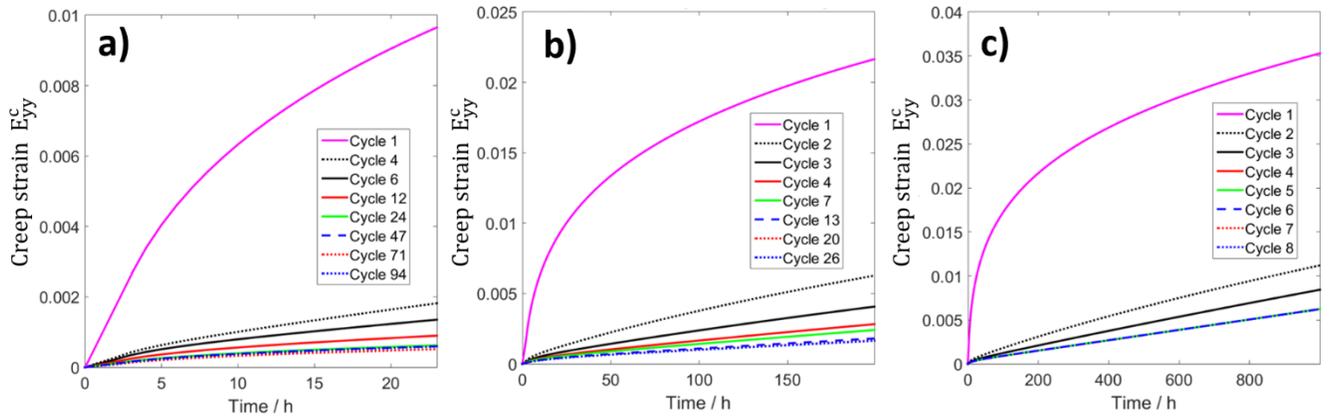

**Figure 12.** Predicted creep strain accumulation during load-controlled dwells at $\Sigma_{dwell} = 140$ MPa at different cycles ($\Delta E = 0.3\%$ from Figure 13) for a) 24 h, b) 200 h and c) 1000 h.

The results in Table 5 suggest that approach b) with full re-priming overestimates the accumulated creep strain by a factor of 3 for the history of 9×1000-hour dwells across all strain ranges considered. For shorter dwells, approach b) results in even greater degree of conservatism. This result agrees qualitatively with experimental observations on primary creep regeneration in Type 316H at 650°C [26] which confirms that the current assumption of full-repriming in the R5 high-temperature assessment code is overly-conservative. The accumulated creep strain is overestimated by factors of 10 and 7 for the 45×200h history and 14 and 11 for the 375×24h history at $\Delta E = 0.3\%$ and 0.8%, respectively. Comparison between the estimated cumulative creep strain in all three 9000-hour histories with load-controlled dwells by the two assessment approaches and the actual simulated histories is made in Figure 13. The ratio of accumulated creep strain using the practical approaches ($E_{Pract}^{c,tot}$) and the total creep strain during the simulated history by the SCM ($E_{SCM}^{c,tot}$) is assessed against dwell time. The results in Figure 13 illustrate that the discrepancies between the approaches and actual simulated response increase when the assumed dwell duration in the practical approaches is shorter than the actual dwell duration. This is explained as follows. For both approaches, the discrepancy is associated with the balance between thermal recovery and hardening of the dislocation microstructure during the primary creep stage in each creep dwell. For shorter dwells, e.g. 24 hours, a dynamic balance between hardening and thermal recovery is not fully-reached and the softening effect of coarsening of the dislocation network is limited. The hardening during those initial stages of the dwell has, therefore, a more pronounced effect on the microstructure over a given period of time and a greater amount of relative hardening of the material during the dwell is induced. This causes a further decrease in the creep rate during subsequent dwells, in addition to that resulting from the cyclic hardening outside the dwell. Hence, the actual dwell response will become progressively "harder" than the reference creep response during the first-cycle dwell. However, the overall creep rate during such shorter dwells is still greater than that assumed during a single dwell using approach a). Experimental studies to further verify the predicted behaviour described in this section are needed.

**Table 5.** Creep strain accumulation for $\Delta E$ = 0.3-0.8% after 9×1000-hour, 45×200-hour and 375×24-hour stress dwells and comparison with approaches a) and b) for the three different 9000h histories.

| $\Delta E$ / $\Sigma_{dwell}$ | 9000-hour dwell (a) | 9×1000-hour | | 45×200-hour | | 375×24-hour | |
|---|---|---|---|---|---|---|---|
| | | Approach (b) | Actual history (SCM) | Approach (b) | Actual history (SCM) | Approach (b) | Actual history (SCM) |
| 0.8%, 167 $MPa$ | 0.157 | 0.456 | 0.172 | 1.61 | 0.24 | 7.19 | 0.69 |
| 0.4%, 150 $MPa$ | 0.109 | 0.370 | 0.116 | 1.16 | 0.14 | 4.74 | 0.41 |
| 0.3%, 140 $MPa$ | 0.087 | 0.317 | 0.092 | 0.97 | 0.10 | 3.68 | 0.25 |

The normalized parameter $f_d$, introduced in Section 3.2.1, is evaluated for the load-controlled creep dwell history examined here. During the creep-fatigue history with displacement-controlled dwells (Section 3.2.1) stress relaxation during the dwell causes a continuous decrease in stress and, respectively, a decrease in creep strain rate during the dwell. By contrast, load-controlled dwells result in the accumulation of greater inelastic strains during the cycle since the macroscopic stress is constant. This leads to more pronounced hardening induced in the microstructure during the load-controlled dwells. In turn, this causes a greater decrease in the primary creep rate during dwells in subsequent cycles, as compared to the reference primary creep rate during the first-cycle dwell. This is illustrated in Figure 14, where for load-controlled dwells the value of $f_d$ at $N/N_s \sim 0.5$ is of the order of 0.1-0.3. These values are lower than those estimated for the displacement-controlled dwells in Figure 8 by a factor of ~3. In effect, the simulation results suggest that the estimated extent of creep damage during load-controlled dwells would be overestimated by a factor of 3-10 using approach b). This observation further highlights the conservatism of approach b) in estimation of accumulated creep damage during load-controlled dwells when the assessment procedure assumes that primary creep is fully regenerated in every dwell.

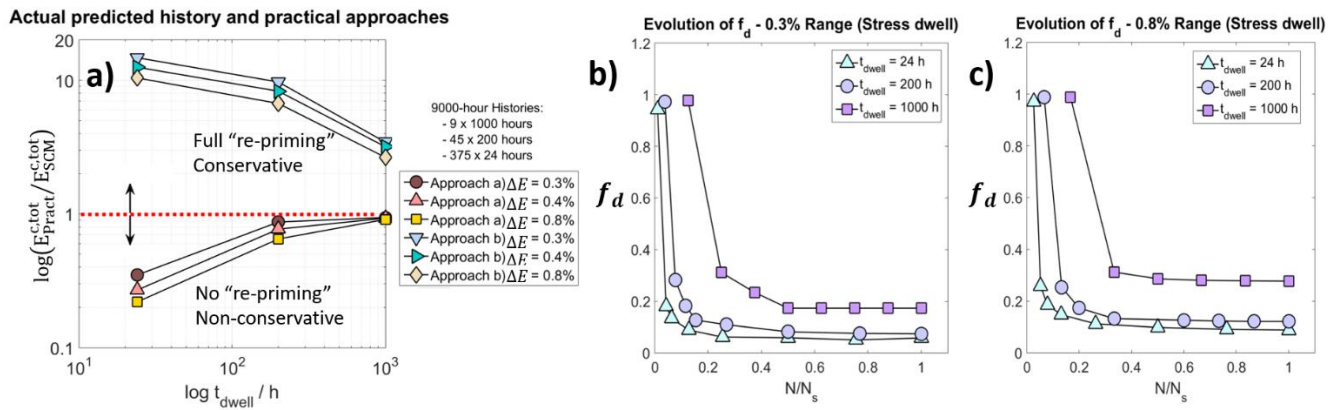

**Figure 13.** a) Discrepancies between practical approaches Section 3.2.2 (a/b) and simulated histories with different load-controlled $t_{dwell}$ and $\Delta E$. In b) and c), evolution of $f_d$ with $\Delta E$, normalised cycles to saturation and dwell times is shown for load-controlled dwells at maximum tensile strain and constant strain range load reversal.

# 4. DISCUSSION

**4.1 EFFECTS OF MICROSTRUCTURAL EVOLUTION ON CREEP-CYCLIC RESPONSE**

The simulation results confirm the effect of prior cyclic hardening on reducing the extent of the primary creep stage in austenitic stainless steel. This is due to the evolution of the dislocation network in the material, as described by Ajaja and Ardell [5] and Joseph et al. [4]. During cyclic loading, the dislocation network is refined due to the reduction in spacing between the obstacles to dislocation motion [11], stemming from the continuous interaction between self/latent hardening and dynamic recovery. Once the dislocation obstacle network is refined, the balance between hardening and thermal recovery during creep loading is achieved quicker. As a result, lower amounts of creep strain are accumulated during primary creep, since the rate of static recovery by dislocation climb would be greater for the higher dislocation densities, as suggested by Morris and Harris [30]. In a similar fashion, the minimum creep rate during the dwell in fully-hardened material is decreased, which stems from the higher dislocation obstacle densities in cyclically-deformed samples. Hence, a greater number density of obstacles is provided which hinders dislocation glide and results in lower inelastic strain rates, agreeing with [4,25]. In addition to the microstructural evolution, it is demonstrated in Section 3.1.2 that the accumulation of creep strain during the primary stage is also dependent on the redistribution of residual stresses between grains. Model predictions illustrate the reduced extent of primary creep in dwells for which the prior deformation history has already established a suitable residual stress field for compatible plastic deformation of the aggregate [14]. The observed phenomenon is more pronounced for larger strain ranges and its effect in plant-operating conditions, where strain ranges are <0.3%, appears to be negligible. As discussed, the model predictions in Section 3.1 are in good qualitative agreement with experimental results from the literature. However, a more systematic experimental study to examine the effects of loading from different stress states on the accumulation of inelastic strain during creep dwells is necessary to confirm fully the findings in Section 3.1.2.

The decrease in saturated stress amplitude of the hysteresis loop during cyclic histories interrupted by displacement-controlled dwells, with the extent of the reduction in $\Sigma_{sat}$ increasing with dwell time, is believed to be as a result of thermal recovery during creep. Thermal recovery coarsens the dislocation network and softens the material. This explanation is plausible since at shorter dwell times, e.g. a 24-hour mid-strain dwell at $\Delta E = 0.3\%$ (see Figure 9.a.), there is a tendency for a more pronounced hardening effect during the stress increase, following completion of the dwell. This suggests that hardening during the primary creep stage of the dwell dominates over thermal recovery for short dwell times. The coarsening of the dislocation network during the dwell is also responsible for the faster saturation of the hysteresis loop with increasing number of cycles. Thermal recovery contributes to the increase in dislocation junction spacing which allows for a quicker balance between hardening and dynamic recovery during cyclic loading. This also explains why when dwells are introduced at mid-strain positions, as compared to the response when dwells are at maximum strain, saturation of the hysteresis loop is reached after a greater number of cycles. Dwells at mid-strain positions are carried out at lower stresses where the relative coarsening of the dislocation network is less pronounced.

The modelling results in Figure 13 suggest that the difference between accumulated creep strain in the history i) estimated by both practical approaches and ii) that predicted by the SCM is sensitive to dwell duration. Nevertheless, strain range also plays a role. At larger strain ranges the start-of-dwell stress is greater. This makes the extent of primary creep during subsequent dwells more pronounced and less affected by the cyclic hardening (increase in dislocation obstacle junction density), prior to the dwell. Hence, the creep behaviour during subsequent dwells will resemble more closely the creep response of the reference dwell in the first cycle. A similar trend in the dwell creep response is observed for the deformation history with displacement-controlled dwells in Section 3.2.1. However, consider an assessment approach based on approach b) in Section 3.2.2, where the effect of cyclic deformation on the dwell is neglected and a dwell on a virgin sample at the same start-of-dwell stress and duration is used to quantify the creep strain. Then, the variation of $f_d$ in Figure 8 suggests that such an assessment approach would result in a more sensible estimation of accumulated creep strain during displacement-controlled dwells, than approach a). Note that at larger strain ranges, the value $f_d$ is over 0.85 for all dwell durations at $\Delta E = 0.8\%$ for displacement-controlled dwells. At large strain ranges, each displacement-controlled creep dwell appears to resemble closely the creep response during the equivalent relaxation dwell with the same duration and start-of-dwell stress without any prior deformation history. This is due to the persistence of the rapid stress-drop at higher start-of-dwell stresses during a relaxation dwell, as observed in [31]. The observations in Section 3.2.2 demonstrate the sensitivity of creep strain estimation during creep-cyclic histories on assessment assumptions and reveal important aspects of the role of strain range and dwell length on the predicted outcome. Experimental data tends to be high $\Delta E$ and short dwell duration, whereas plant operation tends to be low $\Delta E$ and long dwell duration. Therefore, systematic employment of the SCM in evaluation of cyclic-creep histories could ensure that simplified empirical assumptions in assessment approaches remain within reasonable extrapolation bounds. This can also allow for intelligent design of experimental studies to verify the predicted behaviour.

### 4.2 PREDICTIVE CAPABILITIES AND LIMITATIONS OF THE MULTI-SCALE SCM

This article demonstrates various ways in which a multi-scale micromechanical SCM can be used to inform the existing structural assessment procedures, typically used for reactor component lifetime estimation. Findings could also serve to support structural design approaches for next generation high-temperature power plants. Particular emphasis here is placed on providing more insights into the creep-fatigue interactions in austenitic Type 316H stainless steel under AGR plant-relevant thermo-mechanical histories. The model can be used in a systematic framework to evaluate the changing creep properties of the material due to prior cyclic hardening, as described in Section 3.2. It also provides information on the evolution of inelastic strain accumulation during creep or relaxation dwells with increasing cycles (e.g. the parameter $f_d$), as well as the softening behaviour of the hysteresis loop due to the presence of such dwells. Of great relevance to plant assessment are the guidelines provided by the multi-scale model on the variation of hysteresis loop softening behaviour with dwell position. Systematic assessment of saturated stress decrease with dwell time for dwells at different positions can yield an extrapolation framework, which captures the variation of the softening response with dwell time and position. This could potentially be incorporated in the R66 advice. One such practical approach is illustrated in Figure 14 using model predictions from Section 3.2.1. Application of this approach can also support the evaluation of decreasing creep-fatigue life with increasing creep dwell times in advanced austenitic stainless steels, e.g. as reported for Alloy 709 in [32].

The constitutive model, employed in the SCM framework, captures the essential physics of the micromechanical deformation process in polycrystalline Type 316H stainless steel [17,33]. However, it is important to note that some phenomena which could influence the deformation response of this material at very low stresses (e.g. creep strain accumulation due to bypassing of intragranular particles by local climb, diffusive creep) and cyclic deformation at temperatures above 600°C (e.g. change in hardening response due to subgrain/dislocation cell formation) are currently not considered. Further experimental studies are required to evaluate the response of the material at low-stress creep regimes and high-temperature cyclic loading at different strain ranges. These can then guide the enhancement of the existing constitutive model to include potential missing physical phenomena.

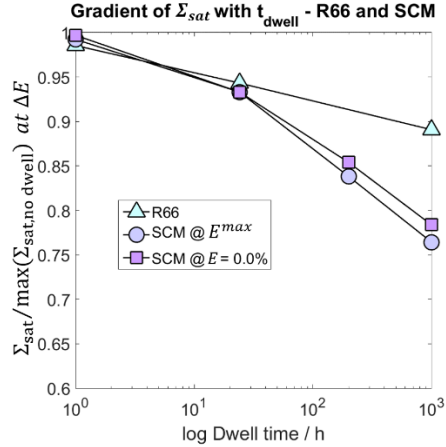

**Figure 14.** Comparison of the predicted decrease in saturated stress amplitude via the R66 data handbook and SCM with dwell time for different dwell positions with respect to strain. Note: The ratio of $\Sigma_{sat}/\Sigma_{sat,no\ dwell}$ across different strain ranges is approx. constant.

Furthermore, in realistic plant conditions many components experience multi-axial loading conditions. It is worth noting that the deformation histories examined in the present study comprise of uniaxially-loaded specimens. The developed SCM is a three-dimensional modelling framework, which could be used to simulate multiaxiality during deformation and inform assessment procedures. As described in [34], the present SCM offers similar predictive capabilities to FE approaches and can study multiaxial loading scenarios when only global deformation of the aggregate is considered. This is at a much lower computational cost and shorter simulation times. In addition to this, both multiaxiality and prior hardening in austenitic stainless steels are known to have a significant effect on the creep failure ductility of the material [35,36]. In its present form, a major limitation of the SCM is its inability to simulate damage and developments are underway to incorporate creep damage into the framework.

## 5. CONCLUSIONS

We have used the self-consistent model (SCM) developed by Hu et al. [11,12,14] and extended by Petkov et al. [15] to study inelastic deformation under cyclic loading histories, typical of those in the boiler section of high-temperature Advanced Gas-Cooled Reactors. We simulate the material response under cyclic loading histories, which contain displacement- and load-controlled dwells, for which macroscopic experimental data is available.

The major conclusions from this study are:

- The multi-scale SCM is a versatile modelling tool that can be used to inform simpler models for structural assessment or design used by industry for cyclic-creep deformation histories where limited or no experimental data is available, e.g. creep dwell times in excess of 200 hours.
- Modelling results demonstrate more softening of the hysteresis loop for dwells longer than 24 hours (N.B. beyond current experimental data) as compared to the guidance provided via conservative empirical advice in the UK's R5 structural assessment procedure.
- The increased extent of softening, predicted by the model, due to the presence of creep dwells reduces significantly the start-of-dwell stress and predicted amount of creep damage generated during the dwell using a ductility exhaustion approach. For a maximum tensile strain displacement-controlled dwell of 1000 hours at a strain range of 0.3%, the SCM predicts a decrease in stress amplitude of approx. 75 MPa, as compared to the stress amplitude for pure fatigue loading. The advice in the R66 data handbook for the same scenario predicts a decrease of 25 MPa. The predicted creep life of the material can vary by a factor of ~10 due to such a difference in stress level.
- The model could supplement the advice in the R66 data handbook by including the effect of displacement-controlled dwell position on the softening response of the material. Systematic assessment of the decrease in saturated stress with dwell time for dwells at different positions can yield numerical factors, capturing the change in softening behaviour.
- The parameter $f_d$, estimated by SCM predictions, can be used directly in assessments, allowing for more accurate estimation of the creep strain generated in dwells during creep-fatigue loading.
- Model predictions suggest that approaches for estimation of creep strain accumulation during dwells, which consider that cyclic deformation has no influence on the creep strain evolution during the dwell, would result in sensible predictions at larger strain ranges. The accuracy of this approach increases when the duration of the assumed dwell during the first cycle resembles more closely the actual duration of the dwells in the history.

## ACKNOWLEDGEMENTS

M.P. Petkov and A.C.F. Cocks are grateful to EDF Energy for supplying the experimental data and supporting the research (Grant R48427/CN001), described in this paper. The authors would like to thank Mike Spindler (EDF Energy) for the useful discussions on evaluating the experimental data and modelling results.

# APPENDIX. SUMMARY OF THE MULTI-SCALE SELF-CONSISTENT MODEL

The micromechanical part of the model captures the contributions to the internal resistance $\tau_{cr}$ of the three dominant obstacles to dislocation motion, i.e. dislocation junctions ($\tau_d$) with mean spacing $L_{di}$; and precipitates ($\tau_p$) with spacing $L_p$, and discrete solutes ($\tau_s$) with spacing $L_s$ in the matrix:

$$\tau_{cr} = \left( \sqrt{\left( \alpha_d G b L_{di}^{-1} \right)^2 + \left( \alpha_p G b L_p^{-1} \right)^2} \right) + \alpha_s G b L_s^{-1} \quad (A.1)$$

where $G$ is the shear modulus, $b$ is the Burger's vector, and $\alpha_p$, $\alpha_d$ and $\alpha_s$ are dimensionless obstacle strength parameters, see [37]. $L_{di}$ is a function of the dislocation junction obstacle density evolution ($\Delta N_d = 1/\Delta L_d^2$) due to self- and latent-hardening as a result of dislocation multiplication related to the increment in the resolved shear strain $\Delta\gamma$ on a slip plane:

$$\Delta N_{di}^{self} = j_s \Delta\gamma_i; \quad \Delta N_{di}^{latent} = j \sum_{k \neq i} \Delta\gamma_k \quad (A.2; A.3)$$

where $j_s$ and $j$ are the hardening fitting parameters and $i = 1\sim 4$, $\Delta\gamma_i$ is the sum of resolved shear strain increments on all three slip systems along the $i$-th slip plane [11]. Dynamic recovery is captured through a model, introduced by Kocks [38], which is applied at the slip system level [15]. Dynamic recovery is a function of the probability that a dislocation segment length ($\Delta L_r$) is annihilated for given $\Delta\gamma$:

$$\Delta N_{di} = -\Delta L_r N_{di} \Delta\gamma_i / b \quad (A.4)$$

The thermal recovery of the dislocation forest is modelled by assuming that at elevated temperatures long dislocation links grow at the expense of shorter ones, as described by Hu and Cocks [14]. The decrease in dislocation junction number density due to static recovery can be expressed as

$$\Delta N_{di} = -2 W_c D_c G b^5 N_{di}^3 \Delta t / kT \quad (A.5)$$

where $D_c$ is the core diffusivity and $N_{di}$ is the dislocation junction obstacle number density on the $i$th slip plane [17], and $W_c$ is the free parameter controlling static recovery of the dislocation structure. The deformation of the anisotropic grains is modelled using crystal plasticity (CP) theory [11]. The plastic strain rate in each crystal is determined by summing the slip rate, $\dot\gamma$, on each individual slip system, $\alpha$, due to thermally-activated dislocation glide as proposed in [39] according to

$$\dot\gamma^{(\alpha)} = \dot\gamma_0 \exp\left( -\frac{\Delta F_0}{kT} \left( 1 - \left| \left( \tau^{(\alpha)} + \bar\tau_m^r \right) / \tau_{cr}^{(\alpha)} \right|^{3/4} \right)^{4/3} \right) \text{sgn}(\tau^{(\alpha)} + \bar\tau_m^r) \quad (A.6)$$

where $\Delta F_0$ ($\alpha_0 G_0 b^3$) is the required free activation energy, $k$ is Boltzmann constant, $\bar\tau_m^r$ is the Type III (micro-) internal residual stress [13], $\dot\gamma_0$ the reference shear strain rate. The parameter $\alpha_0$ which controls the overall effectiveness of obstacles in Eq. A.6 is an additional free parameter here within the physical range, described in [20]. The polycrystalline aggregate is modelled via a self-consistent scheme [11], similar to that proposed in [40] and [41] (KBW model) by treating grains as anisotropic inclusions in an elastic matrix. The SC scheme allows the increment in residual intragranular stress, $\Delta X^a$, to be obtained

$$\Delta X^a = L^i \left( S^i - I \right) \left[ C^a L^i \left( I - S^i \right) + S^i \right]^{-1} \Delta\varepsilon^{ta} \quad (A.7)$$

with $S^i$ being the Eshelby tensor, $C^a$ is the anisotropic grain compliance matrix and $L^i$ is the stiffness of the isotropic matrix surrounding the grains. The incompatible plastic mismatch strain accumulated between grains, $\Delta\varepsilon^{ta}$, is obtained after the KBW model according to $\Delta\varepsilon^{ta} = \Delta\varepsilon^p - T_{ori}^{-T} \Delta E^p$ [17].